Discovery notes

Open Access

# Transduplication resulted in the incorporation of two protein-coding sequences into the *Turmoil*-1 transposable element of *C. elegans*

Noa Sela[1], Adi Stern[2], Wojciech Makalowski[3], Tal Pupko[2] and Gil Ast*[1]

Address: [1]Department of Human Molecular Genetics and Biochemistry, Sackler Faculty of Medicine, Tel Aviv University, Tel Aviv 69978, Israel, [2]Department of Cell Research and Immunology, George S. Wise Faculty of Life Sciences, Tel Aviv University, Tel Aviv 69978, Israel and [3]Institute of Bioinformatics, Faculty of Medicine, University of Muenster, Muenster D-48149, Germany

Email: Noa Sela - noasela@post.tau.ac.il; Adi Stern - sternadi@post.tau.ac.il; Wojciech Makalowski - wojmak@uni-muenster.de; Tal Pupko - talp@tauex.tau.ac.il; Gil Ast* - gilast@post.tau.ac.il

* Corresponding author

## Abstract

**:** Transposable elements may acquire unrelated gene fragments into their sequences in a process called transduplication. Transduplication of protein-coding genes is common in plants, but is unknown of in animals. Here, we report that the *Turmoil*-1 transposable element in *C. elegans* has incorporated two protein-coding sequences into its inverted terminal repeat (ITR) sequences. The ITRs of *Turmoil*-1 contain a conserved RNA recognition motif (RRM) that originated from the *rsp-2* gene and a fragment from the protein-coding region of the *cpg-3* gene. We further report that an open reading frame specific to *C. elegans* may have been created as a result of a *Turmoil*-1 insertion. Mutations at the 5' splice site of this open reading frame may have reactivated the transduplicated RRM motif.

**Reviewers:** This article was reviewed by Dan Graur and William Martin. For the full reviews, please go to the Reviewers' Reports section.

## Findings

The possible contribution of transposable elements to the proteome has been discussed in several publications [1-7] and has provoked much debate [8]. Many mechanisms are known to increase the protein-domain repertoire, e.g., domain duplication, substitution mutations, insertions, deletions, and domain rearrangements [9]. In metazoans, a transposable element may result in transduction, in which a DNA segment downstream of transposable elements is mobilized as part of an aberrant transposition. This may result in gene duplication or exon shuffling, subsequently enriching the protein repertoire [10-13]. However, in the process of transduction, the transposable element does not acquire gene fragments as part of its sequence.

In plants, on the other hand, thousands of transposable elements contain duplicated gene fragments, captured in a process termed transduplication. Transduplication is a potentially rich source of novel coding sequences within rice and *Arabidopsis thaliana* [14-16]. Recently, transduplications of small nucleolar (sno) RNA by retroposon-like non-LTR transposable elements were found in the *C. elegans* [17] and platypus genomes [18].

The *Harbinger* superfamily of "cut-and-paste" DNA transposons was discovered through *in silico* studies [19]. This superfamily is characterized by *Harbinger*-specific transposases that are distantly related to the transposases encoded by the IS5-like group of bacterial transposons, such as IS5, IS112, and ISL2. *Harbinger* transposons are not as widespread as the eukaryotic *hAT* and *mariner/Tc1* transposons; they are found in plants and nematodes [19-23] but not in mammals. Usually, *Harbinger* transposons are flanked by 3-bp target site duplications and 25- to 50-bp inverted terminal repeats (ITRs).

*Turmoil*-1 is a 5,024-bp long DNA transposon with 760-bp long ITRs and a *Harbinger*-specific transposase (Figure 1A). These ITRs are unique to *Turmoil*-1 and are not found in other members of the Harbinger superfamily of DNA transposons [24]. One complete copy of the *Turmoil*-1 was found on chromosome II of *C. elegans*; eight *Turmoil*-1 fragments exist in the genome (for detailed information, see Table 1). An analysis of *C. elegans* transposable elements [25,26] revealed that a 205-bp ITR sequence within *Turmoil*-1 is highly similar to a region of two exons separated by an intron of the *rsp-2* gene (see pairwise alignment using bl2seq [27] Figure 1B). These two exons encode the RNA Recognition Motif (RRM), which is found in many eukaryal and bacterial proteins. Specifically, the type of RRM domain present in the *rsp-2* (called RRM1) gene is highly conserved evolutionarily [28]. The high similarity between the ITR sequence and the fragment of the *rsp-2* gene implies that one originated from the other. The antiquity of this domain and a phylogenetic analysis (Figure 2) indicate that *Turmoil*-1 has recently acquired a portion of the *rsp-2* gene sequence into its ITR. Tree reconstruction was performed with the PhyML program version 2.4.5 [29] using among-site rate variation with four discrete rate categories, and the JTT model [30] of sequence evolution.

A comparative analysis of the *rsp-2* gene and the 205-bp region of the gene found in the ITR sequence, revealed that the *Turmoil*-1 sequence has accumulated several point mutations within the 5' splice site that make it non-functional, whereas the 3' splice site of the intron remains intact (the mutations in the 5' splice site region are marked in red in Figure 1B). Since the RRM domain within the *Turmoil*-1 DNA transposon is not under purifying selection to maintain the reading frame or the functionality of the splice sites, these mutations are not unexpected.

Within the same ITR domain of *Turmoil*-1, and very close to the site of insertion of the RRM domain of *rsp-2* gene, there is evidence of another "DNA kidnapping" event. A 131-bp fragment from the coding region of *C. elegans* gene *cpg-3*, which is unique to nematodes, was inserted into the ITR (Figure 1C). No sequences homologous to *Turmoil*-1 flank the *cpg-3* gene. Thus, similar to the *rsp-2* case, a gene fragment from the *cpg-3* most likely was acquired by *Turmoil*-1, and not vice-versa. As the gene fragments are present on both sides of the ITR, capture may have occurred through non-homologous recombination.

**Table 1: *Turmoil*-1 elements within the *C. elegans* genome**

| # | Chr (*) | Start (**) | End (**) | orientation | Containing RRM domain (***) | Containing *cpg-3* fragment | Length of the element (****) |
|---|---|---|---|---|---|---|---|
| 1 | I | 216923 | 217647 | - | + (1–827) | + | PL |
| 2 | I | 5979173 | 5980185 | + | + (1–1177) | + | PL |
| 3 | I | 5980183 | 5982782 | - | + (1–2771) | + | PL |
| 4 | I | 7245338 | 7245520 | + | - (125–317) | + | PL |
| 5 | I | 13973017 | 13973196 | + | - (4691–4882) | + | PL |
| 6 | II | 3351574 | 3356982 | - | + (1–5024) | + | FL |
| 7 | II | 10883165 | 10884063 | - | - (519–1090) | - | PL |
| 8 | II | 11452621 | 11452867 | - | + (323–568) | - | PL |
| 9 | IV | 2900974 | 2901153 | + | - (125–316) | + | PL |

(*) chromosome number
(**) Start and End according to the coordinates of *C. elegans* genome version ce6 (May 2008).
(***) in brackets the start and end relative to the full sequence of the *Turmoil*-1 found on chromosome II 3351574 – 3356982 (#6).
(****) PL – partial length of *Turmoil*-1/FL – full length of *Turmoil*-1.

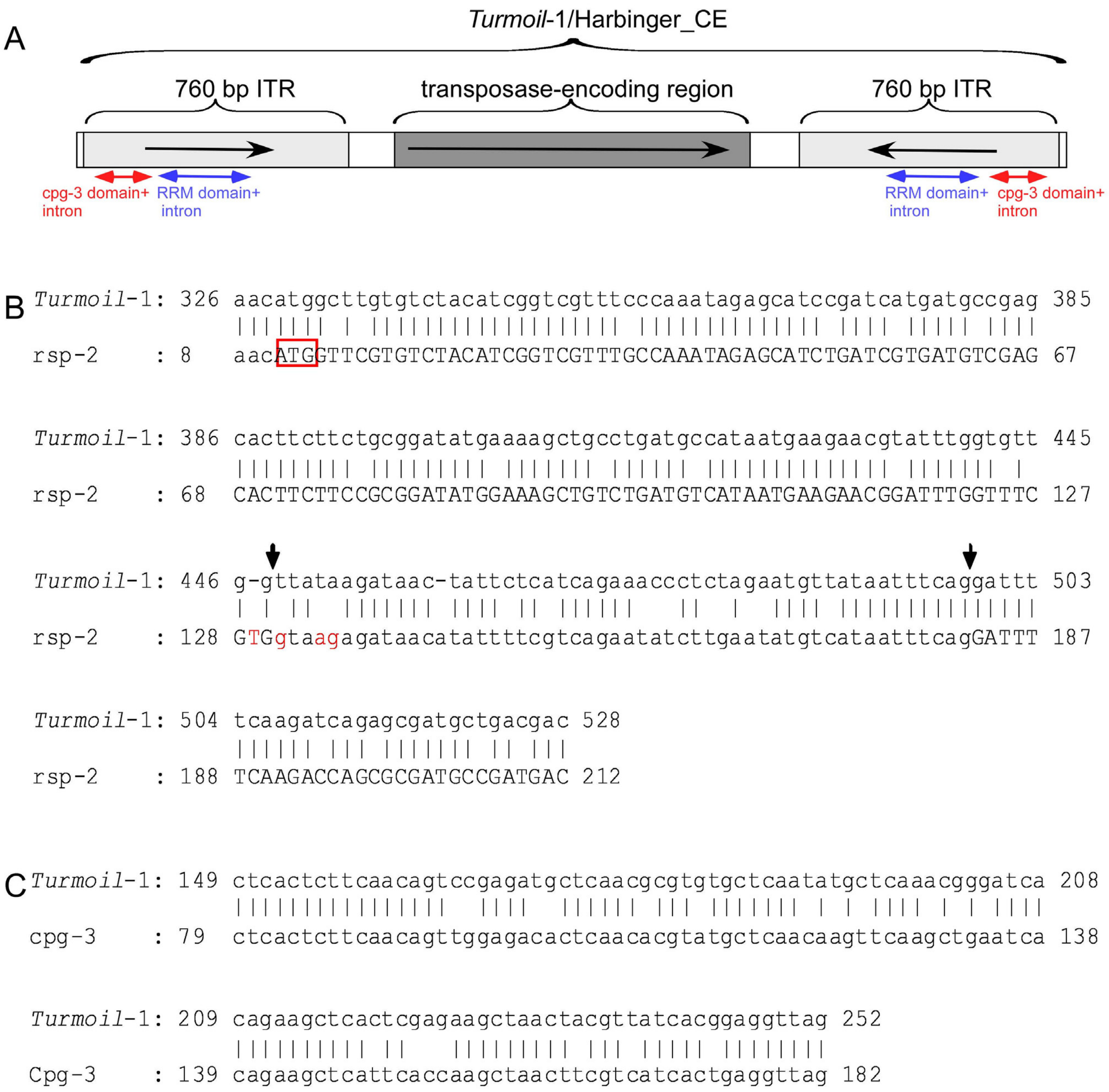

**Figure 1**

**Evidence for two transduplication events within the ITR sequence of the *Turmoil*-1 transposon**. (A) A schematic illustration of the full sequence of *Turmoil*-1 on chromosome II (see table 1, copy number 6) and the structure and sites of inclusion of fragments of the *rsp-2* and *cpg-3* genes within the ITR sequence (B) Pairwise alignment (using bl2seq [27]) between the *Turmoil*-1 sequence and the *rsp-2* sequence. The start codon, which is also the first amino acid of the RRM1 domain in the *rsp-2* gene, is boxed in red. The intron sequence is indicated by lower-case letters; the protein-coding region is indicated by upper-case letters. The positions of the 5' and 3' splice sites (5'ss and 3'ss, respectively) are marked with arrows. The nucleotides in the 5' splice site of the *rsp-2* gene that were mutated in the sequence of the transposable element, thereby abolishing the original 5' splice site, are shown in red. (C) Pairwise alignment between *Turmoil*-1 and the *cpg-3* sequence.

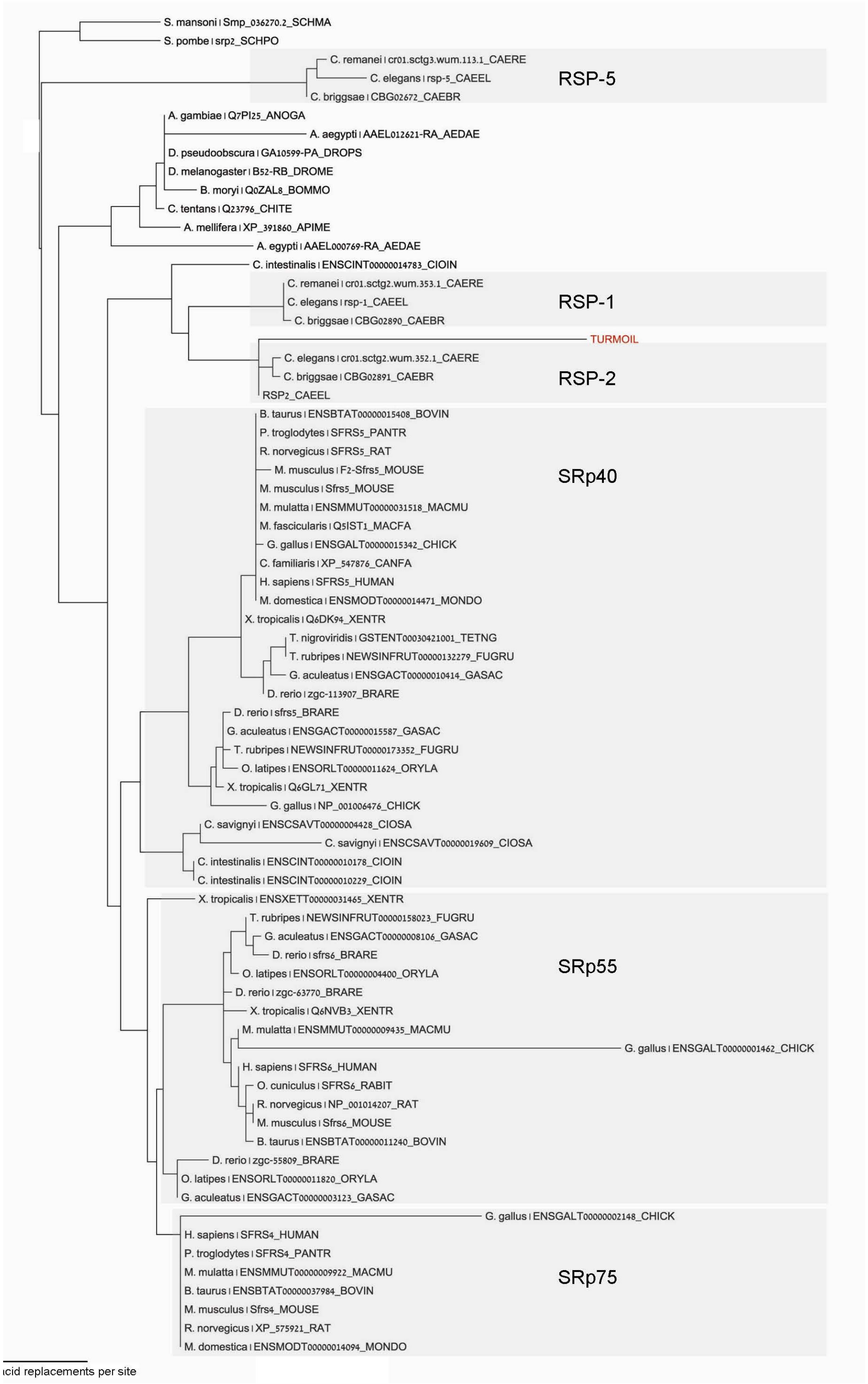

**Figure 2**
**Maximum likelihood tree of the RRM domain**. The RRM domain sequence, which is part of the ITR of *Turmoil*-1, is indicated in red. The tree shows that the RRM domain within *Turmoil*-1 is derived from the ancestral RRM domain rather than vice versa.

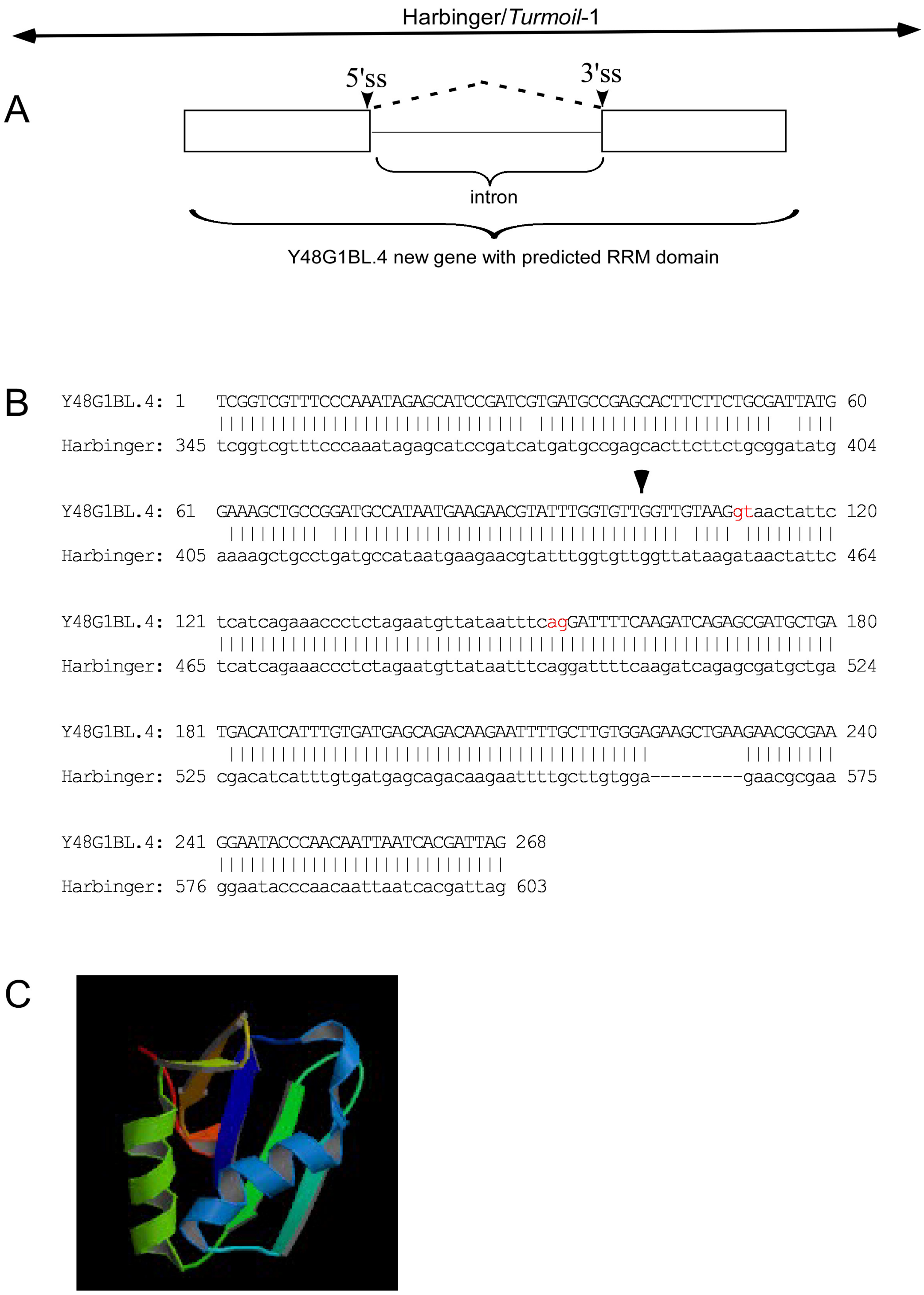

**Figure 3**

***Y48G1BL.4* gene with predicted RRM domain**. (A) A schematic illustration of the *Y48G1BL.4* gene with predicted exons and an intron, and the position of the *Turmoil*-1 sequence within the genome. (B) Pairwise alignment (using bl2seq) between *Turmoil*-1 and the *Y48G1BL.4* DNA sequence. The intron sequence is indicated by lower-case letters and the protein-coding region by upper-case letters. The potential new 5' splice site is marked in red and the 5' splice site position within the original RRM domain is marked with an arrow. (C) The predicted three-dimensional structure of the RRM domain generated from *Y48G1BL.4*.

One of the *Turmoil*-1 copies (number 1 in Table 1) contains within it an open reading frame (ORF) with the accession number *Y48G1BL.4*. It contains two putative exons and an intron (Figure 3), which are similar to the RRM domain. The 5' splice site that corresponds to that in the *rsp-2* transcript has been mutated. A novel 5' splice site is most likely located nine nucleotides downstream from the original one. At this site, a point mutation changed an AT into a GT dinucleotide (marked in red in Figure 3). Usage of this 5' splice site maintains the ORF equivalent to that of the RRM domain of rsp-2 with the exception of the addition of three amino acids. These additional residues should have a negligible effect on the three-dimensional structure of the RRM domain (Figure 3C). This ORF, however, may not be transcriptionally active as its sequence is only found in the UNIPROT database (accession number Q9N3P9), and there is no EST or cDNA supporting evidence. If this is the case, it would be consistent with reports that indicate that all known transduplicates in rice, in spite of their genomic abundance, are pseudogenes [16].

Our analysis indicates that *Turmoil*-1 of *C. elegans* has captured two unrelated coding sequences within its ITRs at proximate locations. The presence of a transduplication "hotspot" in this region may be tentatively inferred. This analysis reveals a transduplication of protein-coding regions in *C. elegans* and strengthens the hypothesis that protein domains may be mobilized by transposable elements.

## Abbreviations

RRM: RNA Recognition motif; ITR: inverted terminal repeat; ORF: open reading frame.

## Competing interests

The authors declare that they have no competing interests.

## Authors' contributions

NS and AS did the experiments and analysis. WM interpreted the results. GA and TP supervised the study. NS, AS, WM, TP and GA wrote the paper.

## Reviewers' comments

### *Reviewer's report 1: Dan Graur, Department of Biology & Biochemistry University of Houston, Texas, USA*

A very straightforward report – I have no other comments.

### *Reviewer's report 2: William Martin, Institut fuer Botanik III, Heinrich-Heine Universitaet Duesseldorf, Germany*

This is an interesting and straightforward paper reporting the presence of transduplication in Caenorhabditis. The report of transduplication in animals would appear to be novel and certainly of sufficient interest to warrant publication. It might be the seed of a larger transduplication avalanche in animals, we'll see. I think the paper is fine for publication with the exception of "open read [ing] frame" in the abstract.

*Author's response*

Thanks for your comment – the typo was corrected.

## Acknowledgements

We thank Prof. Jerzy Jurka for critical reading of the manuscript. This work was supported by the Israeli Ministry of Science and Technology (MOST) and by grants from the Israel Science Foundation (1449/04 and 40/05), MOP Germany-Israel, GIF, ICA through the Ber-Lehmsdorf Memorial Fund, and DIP and EURASNET. AS is a fellow of the Complexity Science Scholarship program and is supported by a fellowship from the Israeli Ministry of Science.